\documentclass[twocolumn,journal]{IEEEtran}
\usepackage{epsfig,epsf,times,amssymb,amsmath}
\title{Design of LDPC Codes using Multipath EMD Strategies and Progressive Edge Growth} 

\author{C. T. Healy and Rodrigo C. de Lamare
\thanks{C. T. Healy and R. C. de Lamare are with  CETUC-PUC-Rio, 22453-900 Rio de Janeiro, Brazil, and
also with the Communications Research Group, Department of
Electronics, University of York, YO10 5DD York, U.K. (e-mail:
rcdl500@ohm.york.ac.uk). This work was supported in part by CNPq and
FAPERJ in Brazil.}}

\begin{document}
\maketitle

\begin{abstract}
Low-density parity-check (LDPC) codes are capable of achieving
excellent performance and provide a useful alternative for high
performance applications. However, at medium to high signal-to-noise
ratios (SNR), an observable error floor arises from the loss of
independence of messages passed under iterative graph-based
decoding. In this paper, the error floor performance of short block
length codes is improved by use of a novel candidate selection
metric in code graph construction. The proposed Multipath EMD
approach avoids harmful structures in the graph by evaluating
certain properties of the cycles which may be introduced in each
edge placement. We present Multipath EMD based designs for several
structured LDPC codes including quasi-cyclic and irregular repeat
accumulate codes. In addition, an extended class of
diversity-achieving codes on the challenging block fading channel is
proposed and considered with the Multipath EMD design. This combined
approach is demonstrated to provide gains in decoder convergence and
error rate performance. A simulation study evaluates the performance
of the proposed and existing state-of-the-art methods.

\end{abstract}
\begin{keywords}
Channel coding, Low-density parity-check codes, Iterative
decoding
\end{keywords}


\section{Introduction}


Low-density parity-check codes \cite{Gallager} are a class of
iteratively decoded capacity-approaching codes. Due to excellent
performance and low-complexity, parallelisable decoding, this class
of codes is increasingly presented as an option for use in wireless
standards, for example DVB-S2, IEEE 802.11 (Wi-Fi) and the IEEE
802.16e standard for WiMAX.

Irregular LDPC codes \cite{Luby} offer improved performance in the
low to medium signal-to-noise ration (SNR) region of operation.
Asymptotic analysis of the threshold behavior of irregular LDPC
codes for a given set of code parameters allows identification of
optimal irregular LDPC ensembles \cite{DE1} and predicts well the
performance of LDPC codes at large block lengths. However, at short
to medium block lengths, closed paths in the graph invalidate the
assumption that messages passed in the iterative decoding are
independent. In practical terms, the closed paths (cycles) in the
graph harm error rate performance and introduce an error floor, a
reduction in error rate performance improvement with improving
channel conditions. At larger block lengths, graphs selected
randomly from the code ensemble with desired parameters generally
perform well, but at shorter lengths care must be taken in graph
selection or construction \cite{Mao_heuristic}-\cite{DOPEG_Comms}.
The trellis-based approach of \cite{Tian} demonstrated that cycles
do not contribute to the error rate uniformly. {A tree-based
approach for graph construction based on progressive edge growth
(PEG) was presented in \cite{PEG_Hu} and was later improved by using
Approximated Cycle Extrinsic Message Degree (ACE) properties
\cite{ace_letters,ace_trans}. PEG-based designs have also been
significantly successful in graph construction for both unstructured
and structured classes
\cite{DOPEG_Comms,Xiao,uchoa2011ldpc,pegbf,PEG_EMD_prev,PEGQC_Li,protograph,uchoa2012generalised,uchoa2013repeat},
including approaches for dealing with stopping sets
\cite{richter,dinoi}. }

In this paper we propose a Multipath EMD strategy for PEG-based
graph construction of LDPC codes which leads to improved error floor
performance in the constructed code realization. The proposed method
is flexible in rate, irregular node degree distributions and the
class of constructed code. It is implemented as a progression of
decision metrics which are used to prune a set of candidate
placements, with the decisions based on an indirect measure of the
impact of each placement on the graph as a whole. The goal is to
reduce the effects of the unavoidable graph structures present at
finite block lengths on the iterative LDPC decoding process.
Following the presentation of the proposed metric, a novel class of
codes capable of approaching the outage limit on block fading
channels with different numbers of fading coefficients is
introduced. These codes are demonstrated to perform excellently at
short block lengths, but require a relatively large number of
decoder iterations to achieve the desired performance. The proposed
Multipath EMD construction is demonstrated to provide considerable
gains in terms of decoder convergence. A detailed justification for
each of the main contributions of the paper, namely the proposed
novel graph construction approach and the proposed
diversity-achieving class of codes, is provided. A simulation study
of the proposed construction along with the existing
state-of-the-art is provided, showing the gains achievable for a
number of structured code classes on the AWGN channel and for the
proposed novel reduced structure diversity-achieving codes on the
block fading channel.

In summary, this paper has the following contributions:
 {
\begin{itemize}
\item The proposed Multipath EMD graph construction strategy.
\item The proposed code class design to operate on a block fading channel with an arbitrary number of fading coefficients.
\end{itemize}}


The rest of this paper is laid out as follows: In Section \ref{sec:channels} the channel models considered in this paper are described. In Section \ref{sec:metric} the proposed Multipath metric progression is detailed, including a discussion of the previous approaches, and a mathematical and algorithmic description of the proposed approach.  In section \ref{sec:BF_red_struct}, the novel code class for use on the block fading channel is described, a discussion of prior work for the channel with two fading coefficients motivates the expansion first to the channel with three fading coefficients and then to the general case. A note on the versatile use of these codes on channels with varying numbers of fading coefficients through the use of a simple puncturing scheme is also provided. In Section \ref{sec:results_Ch4}, a detailed simulation study is provided for the work proposed in this paper. Section \ref{sec:summary_Ch4} provides a brief conclusion to the paper. The appendix provides some analysis and discussion of the work proposed in the paper.

%

\section{Channel Models}
\label{sec:channels}

\begin{figure}[!ht]
\centering
\includegraphics[width=0.4\textwidth]{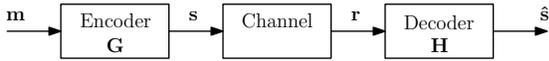}
\caption{A general LDPC coding system}
\label{fig:coding_syst_blockdiagram}
\end{figure}

A general LDPC coding system is considered in this work, as shown in Fig. \ref{fig:coding_syst_blockdiagram}, where a message represented by the $1\times k$ vector $\mathbf{m}$ is encoded to the length $1\times N$ code word vector $\mathbf{s}$, subjected to the channel such that the decoder operates on the vector $\mathbf{r}$ to produce an estimate of the code word $\mathbf{\hat{s}}$.

In this paper a number of channels are considered. The received vector $\mathbf{r}$ is given by
{
\begin{equation}
\label{eqn:received_word}
\mathbf{r} = \left[ \alpha_1 s_1 ,~ \alpha_2 s_2 , \cdots ,\alpha_N s_N \right] + \mathbf{n}.
\end{equation}
}
\noindent For the AWGN channel
{
\begin{equation}
\alpha_1 =  \alpha_2 =   \cdots =  \alpha_N = 1.
\end{equation}
}
\noindent and $\mathbf{n}$ is the vector of Gaussian noise samples
{
\begin{equation}
\mathbf{n} =  \left[ n_1,~ n_2, \cdots, n_N\right],
\end{equation}
}
\noindent where $n_i \sim \mathcal{N}(0,\sigma^2)$.

For the block fading channel with F independent fades,
{
\begin{eqnarray}
\label{eqn:fading_coefs}
\alpha_1 =  \alpha_2 =  \cdots =  \alpha_{\frac{N}{F}} = \beta_1, \nonumber \\
\alpha_{\frac{N}{F}+1} =  \alpha_{\frac{N}{F}+2} =  \cdots =   \alpha_{\frac{2N}{F}} = \beta_2, \nonumber \\
\vdots \nonumber \\
\alpha_{N -\frac{N}{F}+1} =  \alpha_{N - \frac{N}{F}+2} =  \cdots =   \alpha_{N} = \beta_F,
\end{eqnarray}
}
\noindent where the fading coefficients are Rayleigh distributed $\beta_j \in \mathbb{R}^{+}$ and again the noise is Gaussian, $n_i \sim \mathcal{N}(0,\sigma^2)$.

For the fast fading channel, each coded bit is subjected to independent fading coefficients which are Rayleigh distributed $\alpha_i \in \mathbb{R}^{+}$. This is equivalent to the block fading channel with $F = N$ fades, and the additive white Gaussian noise samples are given by $n_i \sim \mathcal{N}(0,\sigma^2)$.

\section{Proposed Multipath EMD Metric Progression}

\label{sec:metric}

In this section, the basis for the proposed construction algorithm, the novel Multipath EMD metric progression, is introduced and discussed in detail. An overview of previous construction metrics motivates the approach considered in this work. The new metric progression is then outlined in detail, and the pseudocode for the proposed construction is provided, explicitly describing the proposed Multipath EMD construction algorithm.

\subsection{Preliminaries}

\subsubsection{Design Problem}

The design problem for LDPC codes is to find a member of the code ensemble which provides good performance on the channel of interest. Certain related graph structures, namely pseudo-codewords \cite{pseudocodewords}, stopping sets \cite{Di_stoppingset} and trapping sets \cite{Richardson_ErrorFloor} have been shown to be responsible for error events under iterative decoding. However, optimisation of the code graph with respect to these structures directly is in general too complex to be achievable for practical code lengths \cite{McGregor_TrappingSets}. Instead we resort to optimisation of these graph properties indirectly. Short length cycles have long been known to harm performance severely and girth optimisation results in improved performance over randomly constructed graphs. In fact, for the low to medium signal-to-noise ratio range of operation, the powerful progressive edge growth (PEG) algorithm provides among the best performance achievable by improving cycle length alone. Every pseudo-codeword is associated with a stopping set, which are formed from connected cycles. On the binary erasure channel (BEC) stopping sets dictate performance entirely \cite{Di_stoppingset_stoppingset}. Careful graph construction with respect to stopping sets yields improved performance \cite{Tian}\cite{Xiao}.

\subsubsection{Definitions}

In constructing LDPC code graphs, greater connectivity has been demonstrated to influence error floor performance of the graph \cite{Tian}\cite{Xiao}. In the following, a number of basic definitions are provided which will clarify the development of the Multipath EMD metric.

\emph{Definition 1: A cycle is a closed path in a Tanner graph with no repeated edges.}

\emph{Definition 2: A stopping set is a set of variable nodes for
which every check node neighbour of any member of the set is
connected to the set at least twice }\cite{Di_stoppingset}.

This structure leads to an uncorrectable error on the BEC and constitutes a worst-case scenario in terms of independence of messages passed under iterative decoding in general.

\emph{Definition 3: The extrinsic message degree (EMD) of a set of variable nodes (or a cycle) is the number of check node neighbours singly connected to that set (or cycle)} \cite{Tian}.

Clearly, the EMD of a stopping set is zero.

\emph{Definition 4: The approximate cycle EMD (ACE) for a variable node is the degree of the variable node minus two \cite{Tian}.}

The ACE metric provides an approximate measure of the EMD of a cycle by assuming that all check node neighbours which are not directly involved in the cycle are connected to the cycle only once.

\begin{figure}[!ht]
\centering
\includegraphics[width=68mm]{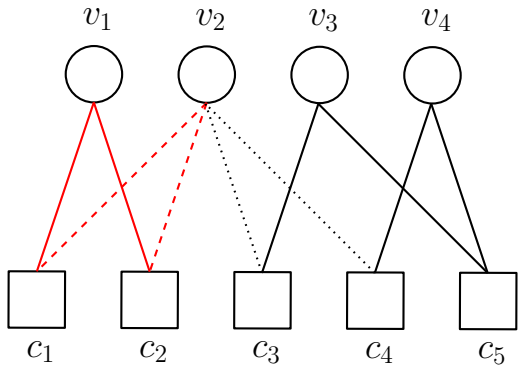}
\caption{Small Tanner Graph with cycles}
\label{fig:Cycle_StopSets}
\end{figure}

Figure \ref{fig:Cycle_StopSets} outlines the points reviewed in
\emph{Definitions 1 - 4}. Two cycles are shown, the length 4 cycle
$[ v_1,c_1,v_2,c_2 ]$ and the length 6 cycle $[
v_2,c_3,v_3,c_5,v_4,c_4 ]$. Neither set $\{v_1, v_2\}$ nor $\{ v_2,
v_3, v_4 \}$ alone is a stopping set, as both have extrinsic
connections from $v_2$, the dotted black lines are extrinsic with
respect to the set $\{v_1, v_2\}$ while the dashed red lines are
extrinsic with respect to the set $\{ v_2, v_3, v_4 \}$. It is clear
that the set $\{ v_1, v_2, v_3, v_4 \}$ is a stopping set, formed
from the combination of the length 4 and length 6 cycles. Note that
using an ACE style metric based on variable node weight, the set $\{
v_1, v_2, v_3, v_4 \}$ would appear to have two extrinsic
connections, however a true EMD calculation shows that this set has
no extrinsic connections and so is a stopping set. 


\subsection{Metric}

\label{subsec:Metric}

The PEG construction algorithm proceeds columnwise and edgewise. The task at each edge placement is to prune the set of all check nodes to a single survivor, which is connected to the variable node under consideration. The PEG algorithm in its original form selected survivors according to the longest path metric, resulting in creation of the longest possible cycle, followed by the minimum current check node weight metric which gives the graph the desirable near-regular check node distribution.

The IPEG algorithm includes a further set-pruning step, based on the path ACE metric which provides an approximate measure of the level of connectivity which the cycle or cycles created will have to the rest of the graph \cite{Xiao}. This connectivity determines the performance of the graph under iterative decoding through its influence on stopping set creation. The performance improvements achieved in the error floor region by the codes constructed by IPEG algorithm support the efficacy of applying graph connectivity and stopping set avoidance principles to graph construction.

Another work in the literature adds a further set pruning step based on the exact EMD measure of a set of variable nodes, with the set being that of all variable nodes contained in all paths between a particular candidate check node and the variable node of interest \cite{PEG_EMD_prev}. The candidate with the largest path set EMD is chosen as the survivor. For the case when there is a single path between the candidate and variable node, this measure gives an exact EMD of the cycle created. However, when multiple paths exist then the EMD measure produced will not reflect the likelihood that the individual cycles created participate in stopping sets, but rather the likelihood that the combination of all those cycles combined will form or participate in a stopping set. This is clearly an issue as smaller stopping sets are much more harmful to performance than large ones, and each individual path constitutes a cycle which may participate in a smaller stopping set, an eventuality which is not reflected by the metric proposed in that work. Nevertheless, the EMD-based metric progression did offer further improvements in error rate performance in the error floor region.

\begin{figure}[ht!]
\centering
\includegraphics[width=0.4\textwidth]{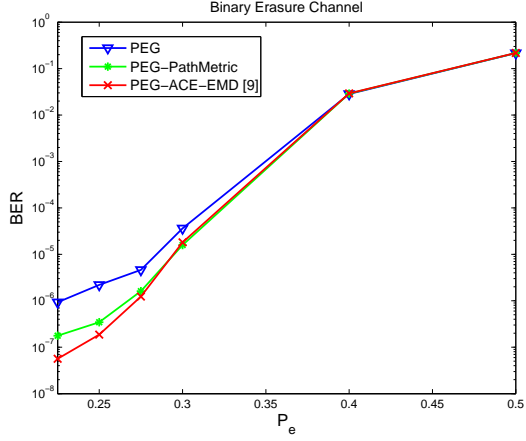}
\caption{Plot showing the performance on the BEC of the graph
constructed with the first stage of the proposed metric progression
only, compared to the codes constructed by the standard PEG
algorithm. } \label{fig:res_BEC1_MinPathsMetric}
\end{figure}

In this work, an alternative progression of metrics is proposed for choosing the survivor check node from the set of candidates. First, the PEG tree expansion is carried out to find the set of check nodes at equal maximum distance from the variable node of interest. This reduces the set of check nodes to be considered greatly and has been demonstrated as one of the best approaches currently known. As with the original algorithm, the minimum node weight metric is also applied, further reducing the set of check nodes to be considered. For each of these survivors, in an operation to be outlined in the following section, for each candidate check node each distinct path from root variable node to candidate check node is identified and the precise EMD of each path is computed. From the current candidate check node set, those check nodes with fewest paths from variable node to check node are selected to survive. The justification for this selection metric lies in the fact that stopping sets are formed either from single zero EMD cycles (comprised of only weight two variable nodes) or from the combination of cycles such that they are joined by all of their respective extrinsic edges. Thus, reducing the number of small cycles in the graph has the effect of reducing the likelihood of stopping set creation. Note that the individual zero EMD cycles are easily avoided in graphs constructed by the PEG algorithm by applying the constraints on the number of weight two variable nodes of \cite{RA_Yang}. Fig. \ref{fig:res_BEC1_MinPathsMetric} provides results confirming the effectiveness of this metric for the BEC channel, comparing the performance of the standard PEG constructed graph with that of the PEG algorithm and minimum path number metric. Finally, for the remaining check nodes which have equal maximum distance, minimum weight and the same minimum number of shortest paths from the variable node of interest, the average EMD of the shortest paths is computed and the candidate with the largest value is chosen for edge placement. This choice of average EMD across all paths rather than the EMD of the path with worst connection is again made to reduce the overall likelihood of stopping set creation in the graph construction. The results presented in Section \ref{sec:results_MEMD} demonstrate the efficacy of avoiding stopping set creation throughout the graph in this manner, with a gain of approximately $0.5$dB observed for the QC-LDPC graph and of approximately $0.25$dB for the IRA graph.

\subsection{Computation of the Metric}

As the metric progression detailed in the following makes use of the notation introduced in \cite{PEG_Hu}, a brief review is useful. The PEG algorithm involves a tree expansion from the root variable node $v_j$, with each level added to the tree including an additional subset of check and variable nodes, up to the level $l$ at which all check nodes are included in the tree, or further expansion adds no new check nodes. The set of check nodes reached at level $l$ is denoted $\mathcal{N}_{v_j}^l$ while those not yet included are denoted $ \overline{ \mathcal N_{v_j}^l}$. We also define the set of variable nodes included in the tree from node $n_i$ to $m$ levels as $\mathcal M_{n_i}^m$. Note that, for variable nodes, $\mathcal M_{v_j}^0$ contains only $v_j$ while for check nodes $\mathcal M_{c_i}^0$ contains the immediate variable node neighbours of $c_i$. We denote $ \mathcal C$ the set of all M check nodes.

\begin{figure}[!ht]
\centering
\includegraphics[width=0.4\textwidth]{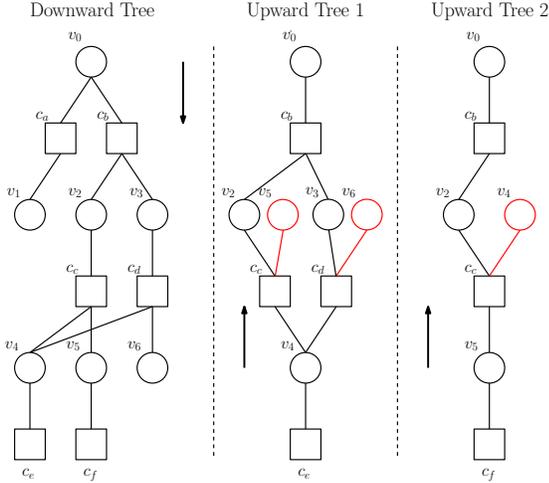}
\caption{The path identification process described by (\ref{eqn:VNs_in_path})-(\ref{eqn:distinct_paths}) as implemented by a comparison of a downward PEG-like tree from the root variable node and an upward tree from each of the candidate check nodes. For a given candidate, any node found at the same level in both the downward and upward tree is contained in the graph between the root variable node $v_0$ and that candidate check node. By (\ref{eqn:s_1})-(\ref{eqn:distinct_paths}) the unique paths are identified.}
\label{fig:PEG_like_trees}
\end{figure}

Once the initial stage of graph construction is complete, the PEG algorithm first returns the subset
{
\begin{equation}
\mathcal{A} = \{ \overline{ \mathcal{N}_{v_j}^{l-1}}: \overline{ \mathcal{N}_{v_j}^{l}} = \emptyset\},
\label{eqn:max_dist}
\end{equation}
}
\noindent and from this set the minimum weight candidates are selected as
{
\begin{equation}
\mathcal{B} = \{ c_i: |\mathcal{M}_{c_i}^0| = \min\limits_{c_x \in \mathcal{A}} |\mathcal{M}_{c_x}^0|  \}.
\label{eqn:min_weight}
\end{equation}
}
\noindent Then for the node pair $\{ v_j, c_i \}$ with $c_i \in \mathcal{B}$ and $L$ levels between $v_j$ and $c_i$, such that $\overline{ \mathcal{N}_{v_j}^{L}} = \emptyset$, the set of variable nodes found at the levels $a$ in all paths between the nodes in this pair is
{
\begin{equation}
\mathcal{D}_a = \mathcal{M}_{v_j}^a \cap \mathcal{M}_{c_i}^{L-a}
\label{eqn:VNs_in_path}
\end{equation}
}
\noindent The sets $\mathcal{D}_a$ must be found for each of the $L$ levels in the graph between $v_j$ and $c_i$. There exists a path between two variable nodes in adjacent levels $a$ and $a+1$ if
{
\begin{equation}
\mathcal{N}_{v_x}^0 \cap \mathcal{N}_{v_y}^{0} \neq \emptyset ~~ , v_x \in \mathcal{D}_a~~,v_y \in \mathcal{D}_{a+1}.
\end{equation}
}
\noindent In order to produce the distinct path number and path EMD metrics, it is necessary to find the set of distinct path variable node sets. These sets are expanded level by level and intialised for the connections from root node to each node in $\mathcal{D}_1$ as
{
\begin{equation}
\label{eqn:s_1}
\mathbf{s}_1 = \{v_j, v_{u_1}\}, \mathbf{s}_2 = \{v_j, v_{u_2}\}, \cdots, \mathbf{s}_{|\mathcal{D}_1|} = \{v_j, v_{u_{|\mathcal{D}_1|}}\},
\end{equation}
}
\noindent because there is an edge connecting the root node $v_j$ to each node in the first level. The number of distinct paths at the first level is $P_1 = |\mathcal{D}_1|$, while the number of distinct paths up to level $a$ is denoted $P_a$. For each path and path variable node set $\mathbf{s}_v$ to level $a$ with $v \in \{1,\cdots, P_a\}$, with variable node $v_a = \mathbf{s}_v \cap \mathcal{D}_a$ the node in $\mathbf{s}_v$ which was found at the current level, there will be $|v_a \cap \mathcal{D}_{a + 1}|$ distinct paths after expanding the set of distinct path sets to level $(a+1)$. The new sets produced from the paths sets to level $a$ and those nodes in level $(a+1)$ are produced according to:
{
\begin{multline}
\label{eqn:distinct_paths}
\mathbf{s}_x = \{ \mathbf{s}_v \cup v_{w_y} : \mathcal{N}_{\mathbf{s}_v \cap \mathcal{D}_{a}}^{0} \cap  \mathcal{N}_{v_{w_y}}^{0} \neq \emptyset  \}, \\
\forall \mathbf{s}_v, v \in \{1, \cdots, P_a\}, \forall v_{w_y} \in \mathcal{D}_{a + 1}.
\end{multline}
}
\noindent Thus a distinct path set for the next level is created for each combination of the path set to the current level $\mathbf{s}_v$ and a node in $\mathcal{D}_{a + 1}$ if there is a path between the node in $\mathbf{s}_v$ at the current level and the node in $\mathcal{D}_{a + 1}$. When this process has been carried out $L-1$ times for the check node $c_i$ then the set of all distinct path sets $\mathbf{S}_{c_i} = \{\mathbf{s}_{p,c_i}\}, p \in \{1, \cdots P_L\}$ to level $L$ is found. The number of distinct paths from $v_j$ to $c_i$, denoted $P_{c_i}$, is the cardinality of the set of all distinct path sets, $P_{c_i} = |\mathbf{S}_{c_i}| = P_L$. The above process must be carried out for each check node in $\mathcal{B}$. The number of distinct paths for each check node is the first element of the proposed metric progression used to prune the set of candidate check nodes:
{
\begin{equation}
\mathcal{C} = \{ c_i: P_{c_i} = \min\limits_{c_y \in \mathcal{B}} P_{c_y} \}.
\label{eqn:CN_pruning}
\end{equation}
}
\noindent In the event that there is a single entry in $\mathcal{C}$ the check node selection procedure terminates and that check node is chosen as the survivor node and the edge $\{v_j, \mathcal{C}\}$ is placed. If, however, there is more than one element in $\mathcal{C}$, the path EMD of each set in $\mathbf{S}_{c_i}$ is computed for $c_i \in \mathcal{C}$. The EMD for the path $p$ connecting to the check node $c_i$ and corresponding to the set $\mathbf{s}_{p,c_i}$ is:
{
\begin{multline}
E_{p,c_i} = |\{c_k: c_k \in \mathcal{N}_{v_b}^0, c_k \not\in \mathcal{N}_{v_c \in \mathbf{s}_{p,c_i}\setminus v_b }^0 \}|, \\
\forall v_b \in \mathbf{s}_{p,c_i}.
\label{eqn:EMD_calc1}
\end{multline}
}
\noindent The EMD $E_{p,c_i}$ for each path can be computed simply by taking the sum of the columns of the parity-check matrix corresponding to the nodes in $\mathbf{s}_{p,c_i}$ and counting the number of $1$s in the resulting vector \cite{PEG_EMD_prev}. For each check node in $\mathcal{C}$, the EMD of (\ref{eqn:EMD_calc1}) is computed for all paths in $\mathbf{S}_{c_i}$ and then the final metric used is computed as the mean of these path EMD values:
{\begin{equation}
\gamma_{c_i} = \frac{1}{P_{c_i}} \sum_{p=1:P_{c_i}}{E_{p,c_i}}.
\label{eqn:EMD_calc2}
\end{equation}
}
\noindent The successful candidate is then the check node with the largest mean path EMD value:
{
\begin{equation}
c_{place} = c_i \in \mathcal{C}:\gamma_{c_i} = \max\limits_{c_z \in \mathcal{C}} \gamma_{c_z}
\label{eqn:EMD_pruning}
\end{equation}
}
Fig. \ref{fig:PEG_like_trees} gives the graphical representation of (\ref{eqn:VNs_in_path})-(\ref{eqn:distinct_paths}), for a particular variable node $v_0$ and two check node candidates labeled $c_e$ and $c_f$, respectively. The tree is expanded to depth two and the nodes at each level for all paths are identified by applying (\ref{eqn:VNs_in_path}) for levels 1 and 2. So, from the downward tree from $v_0$, the variable nodes in the first level of the downward tree are $\mathcal{M}_{v_0}^1 = \{v_1~,v_2,~v_3\}$ while from the first upward tree from $c_e$, it is clear that the nodes reached at level $L-1 = 1$ are $\mathcal{M}_{c_e}^1 = \{v_2~,v_3,~v_5,~v_6\}$, so the nodes which are found at that level in both trees are the nodes present in the graph connecting $v_0$ and $c_e$, $\mathcal{D}_{1} = \{v_2, ~v_3 \}$. The same observation gives $\mathcal{M}_{v_0}^2 = \{v_4~,v_5,~v_6\}$ and $\mathcal{M}_{c_e}^0 = \{v_4\}$ so it is clear that $v_4$ alone appears in the graph from $v_0$ and $c_e$ at this level, $\mathcal{D}_{2} = \{v_4\}$. For the graph between $v_0$ and $c_f$, it is observed that there is a single path only, as $\mathcal{D}_{1} = \{v_2\}$ and $\mathcal{D}_{2} = \{v_5\}$. In this simple example two paths are identified between $v_0$ and $c_e$ while a single path is identified between $v_0$ and $c_f$, and according to the metric progression outlined, $c_f$ would be chosen for the edge placement. In this simple example the EMD calculation and pruning of (\ref{eqn:EMD_calc1})-(\ref{eqn:EMD_pruning}) would not be needed as there is already a single superior check node candidate.

The pseudocode of Algorithm \ref{alg:EMD-PEG} explicitly describes
the algorithm and shows where equations
(\ref{eqn:max_dist})-(\ref{eqn:EMD_pruning}) appear in the structure
of the proposed design algorithm.

\section{Full Diversity Codes with Reduced Structure}
\label{sec:BF_red_struct}

 {In this section a class of codes with fewer
constraints on the graph structure than the Root-LDPC graph
\cite{Boutros_RootLDPC_trans}, } and thus termed reduced structure,
which are capable of achieving the diversity of the block fading
channel is introduced. A Multipath EMD design extension for the
codes with reduced structure for block fading channels is also
presented. The diversity-achieving code class developed in this
section comprises a generalisation of the previously presented code
which achieves the diversity of the channel with $F=2$ only
\cite{Duyck_BF2_unstruct}. In that paper, two results from the
literature were employed:

\emph{For a code to achieve full diversity on the block fading channel, the systematic nodes must be fully recoverable on the block binary erasure channel, i.e. the channel where the fading coefficients take only the values $\beta_j \in [0, \infty]$} \cite{Boutros_RootLDPC_trans}.

And the well-known result concerning stopping sets:

\emph{Under iterative SPA decoding, each uncorrectable error on the
binary erasure channel is associated with a stopping set, stopping
sets fully characterise the error events on that channel.}
\cite{Di_stoppingset}.

Note also that the greatest code rate possible for a code to achieve
the diversity of the channel is $R = \frac{1}{F}$
\cite{Boutros_RootLDPC_trans}.

The rest of this section proceeds as follows: In part
\ref{sec:BF_unstruct_F2}, the previously presented code for the
$F=2$ case is outlined. Part \ref{sec:BF_unstruct_F3} presents the
extension of this approach to the $F=3$ case, while part
\ref{sec:BF_unstruct_F4} indicates the procedure for constructing a
code for a block fading channel with any number of fading
coefficients. Part \ref{sec:BF_unstruct_gain} discusses the coding
gain of the proposed codes.

\label{sec:BF_unstruct}

\subsection{$F=2$ Case}
\label{sec:BF_unstruct_F2}

The work in \cite{Duyck_BF2_unstruct} presented unstructured codes
which achieve full diversity on the block fading channel with $F=2$
given certain constraints on rate, distribution and cycle
properties. To meet the requirement that the systematic nodes be
recoverable on the block binary erasure channel, the fact that
stopping sets fully characterize errors on the binary erasure
channel and thus account for errors on the block erasure channel is
used to produce a new sufficient condition for achieving the
diversity of the channel:

\emph{A systematic node is not recovered if it is a member of a stopping set and if that stopping set is erased}

We term a stopping set containing a systematic variable node a
systematic stopping set. In the $F=2$ case, an uncorrectable error
occurs when all nodes in a systematic stopping set are affected by
the same fading coefficient $\beta_f$.

\begin{figure}[htb]
 \centering
\resizebox{60mm}{!}{
\includegraphics{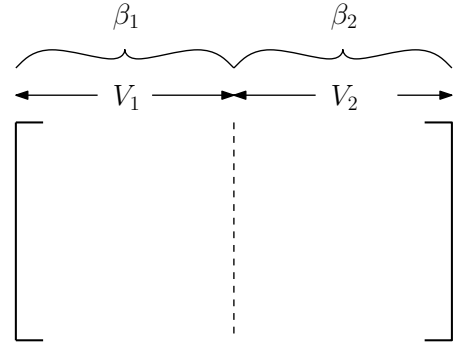}
}
\caption{The rate $ \leq \frac{1}{2}$ code for the block fading channel with $F=2$}
\label{fig:BF2_gen}
\end{figure}

The general parity-check matrix for the code on the $F=2$ channel is
presented in Fig. \ref{fig:BF2_gen}. $V_1$ is the set of variable
nodes affected by $\beta_1$ and $V_2$ is the set of variable nodes
affected by $\beta_2$. All the systematic nodes, $V_{syst}$, are
contained within $V_1$ and protection of these nodes is the goal.
The requirement that the code achieves full diversity on the $F=2$
channel is exactly the requirement that there exists no subset $S
\subseteq V_{syst}$ such that $S$ is a stopping set
\cite{Duyck_BF2_unstruct}. That is: {
\begin{equation}
\exists v_j \in S : \exists c_i, c_i \in \mathcal{N}_{v_j}^0, c_i \notin \mathcal{N}_{v_k \in S\setminus v_j}^0
\label{eqn:BF_stopset_req}
\end{equation}
} \noindent That is, for every subset of the systematic node set,
$V_{syst}$, there exists some variable node with at least one
extrinsic connection with respect to that subset. Then there is no
stopping set contained within $V_{syst}$ and by the previously
stated results of the literature, each node is recoverable on the
block binary erasure channel \cite{Di_stoppingset}, implying that
the code achieves full diversity \cite{Boutros_RootLDPC_trans}.
Thus, the full diversity requirement of the code has been stated as
a constraint on the nature of the code graph.

In \cite{Duyck_BF2_unstruct}, the requirement (\ref{eqn:BF_stopset_req})is achieved by use of the PEG construction and its property concerning cycle creation in the initial construction phase. As no cycle is created in this phase, no stopping set may be created. For weight 2 variable nodes, in the initial graph construction no cycle is created up to the variable node $v_{(M-1)}$ where $M$ is the number of check nodes of the graph \cite{Tian}. This results in the following constraint on code dimension
{
\begin{equation}
K < \frac{N}{2} < (M-1),
\label{eqn:BF_unstruct_requirements}
\end{equation}
}
\noindent which, combined with the specification that the systematic nodes are assigned among these initially constructed weight 2 nodes, leads to a code class which achieves the diversity of the channel.

\subsection{$F=3$ Case}
\label{sec:BF_unstruct_F3}

\begin{figure}[htb]
 \centering
\resizebox{72mm}{!}{
\includegraphics{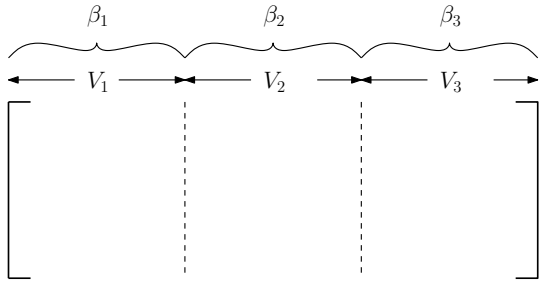}
}
\caption{The rate $ \leq \frac{1}{3}$ code for the block fading channel with $F=3$}
\label{fig:BF3_gen}
\end{figure}

For the channel with $F=3$, the general parity-check matrix is represented in Fig. \ref{fig:BF3_gen}. Again the systematic variable nodes $V_{syst}$ are contained within $V_1$. A stopping set based criterion for full diversity will be developed. In this case it is necessary that the elements of $V_{syst}$ be recoverable on the block binary erasure channel observation where any one of the fading coefficients may be non-zero, or any pair may be non-zero. If all three coefficients are zero ($\beta_1 = \beta_2 = \beta_3 = 0$) the systematic nodes are entirely unrecoverable, and if $\beta_1 =\infty$ the systematic nodes will be fully recovered from the channel irrespective of $\beta_2$ and $\beta_3$. In the case that, if for example, $\beta_3$ is non-zero while $\beta_1=\beta_2=0$, then any stopping set $S \subseteq V_1 \cup V_2$ would be unrecoverable \cite{Di_stoppingset} and likewise for the other single non-zero fading coefficient scenario. Considering only the error rate of the systematic nodes, the necessity that $S$ is not a stopping set is again as expressed in (\ref{eqn:BF_stopset_req}), but the subsets of nodes for which this requirement must hold has expanded to every set where:
{
\begin{equation}
S \cap V_{syst} \neq \emptyset ~:~ S \subseteq V_1 \cup V_2~,~S \subseteq V_1 \cup V_3.
\label{eqn:BF_sets_req_F3}
\end{equation}
}
\noindent This full diversity requirement comprises a constraint on the graphical structure of the code realisation. For the $F=3$ case, the requirement is more difficult to achieve, as there are more configurations of the block erasures which must be considered. However, once a graph is constructed which satisfies (\ref{eqn:BF_stopset_req}) and (\ref{eqn:BF_sets_req_F3}), it is guaranteed to achieve full diversity on the block fading channel with $F=3$, by the results of \cite{Di_stoppingset} and \cite{Boutros_RootLDPC_trans}.

\begin{figure}[!ht]
\centering
\includegraphics[width=0.4\textwidth]{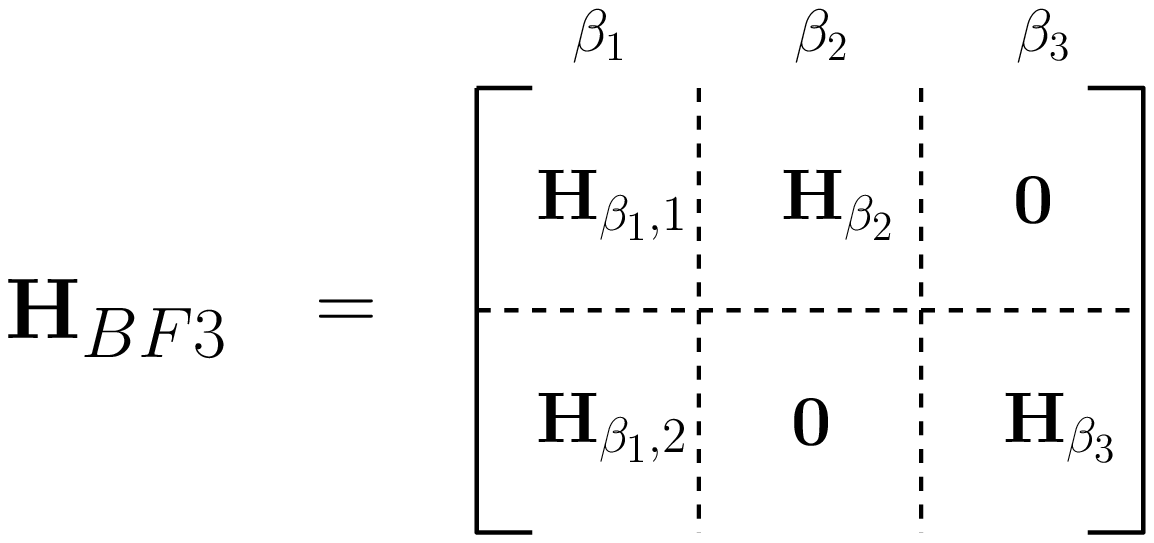}
\caption{Full diversity parity check matrix for the $F=3$ channel}
\label{fig:H_BF3}
\end{figure}

The equations (\ref{eqn:BF_stopset_req}) and (\ref{eqn:BF_sets_req_F3}) together impose the limit that no systematic stopping set exists solely among the variable nodes in $V_1$, among the nodes $[V_1 ~ V_2]$ and among the variable nodes $[V_1 ~ V_3]$. In the Root-LDPC code approach, stopping sets are avoided by the imposition of the root-check structure. However, in order to avoid this structural requirement, an alternative solution is presented in Fig. \ref{fig:H_BF3}. Each of the two subgraphs $[\mathbf{H}_{\beta_1,1} ~ \mathbf{H}_{\beta_2}]$ and $[\mathbf{H}_{\beta_1,2} ~ \mathbf{H}_{\beta_3}]$ are constructed to achieve full diversity on the $F=2$ channel. As such, the subgraph $\mathbf{H}_{\beta_1,1}$ is cycle free, as is $\mathbf{H}_{\beta_1,2}$. Combined, they may have many cycles, however the placement of the null matrices ensures that extrinsic connections exist, to $\mathbf{H}_{\beta_3}$ with respect to $\beta_1, \beta_2$ and to $\mathbf{H}_{\beta_2}$ with respect to $\beta_1, \beta_3$. Thus the systematic variable nodes are recoverable under both $\beta_1 =  \beta_2 = 0,~ \beta_3 = \infty$ and $\beta_1 =  \beta_3 = 0,~ \beta_2 = \infty$. Additionally, under $\beta_1 = 0,~ \beta_2 = \beta_3 = \infty$ the extrinsic connections ensure no systematic stopping sets among the subset of variable nodes affected by $\beta_1$ only. This code therefore completely recovers the systematic bits on the block erasure channel and so achieves full diversity on the block fading channel.

\subsection{Cases with $F>3$}
\label{sec:BF_unstruct_F4}

\begin{figure}[htb]
 \centering
\resizebox{88mm}{!}{
\includegraphics{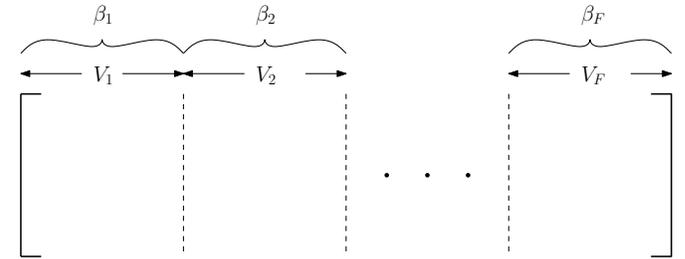}
}
\caption{The rate $ \leq \frac{1}{F}$ code for the general block fading channel}
\label{fig:BFGen_gen}
\end{figure}

In the general case with $F$ fading coefficients, to recover the systematic nodes contained in $V_1$, the stopping set requirement generalises to involve all subsets including elements of $V_1$ and excluding all elements of one or more other fade-affected sets of nodes. Now (\ref{eqn:BF_stopset_req}) must hold for all the subsets described by:
{
\begin{equation}
S \cap V_{syst} \neq \emptyset,
\label{BF_sets_req_F4a}
\end{equation}
}
where
{
\begin{equation}
S \subseteq V_1 \cup V_{k_1} \cup V_{k_2} \cdots \cup V_{k_m}~:~\{k_1\cdots k_m\} \subset \{2,\cdots ,F\}.
\label{BF_sets_req_F4b}
\end{equation}
}
The constraints on the code graph described by Eqns. (\ref{eqn:BF_stopset_req}), (\ref{BF_sets_req_F4a}) and (\ref{BF_sets_req_F4b}) provide a graphical interpretation of the requirements to achieve full diversity on the block fading channel with general $F$.

\begin{figure}[!ht]
\centering
\includegraphics[width=0.4\textwidth]{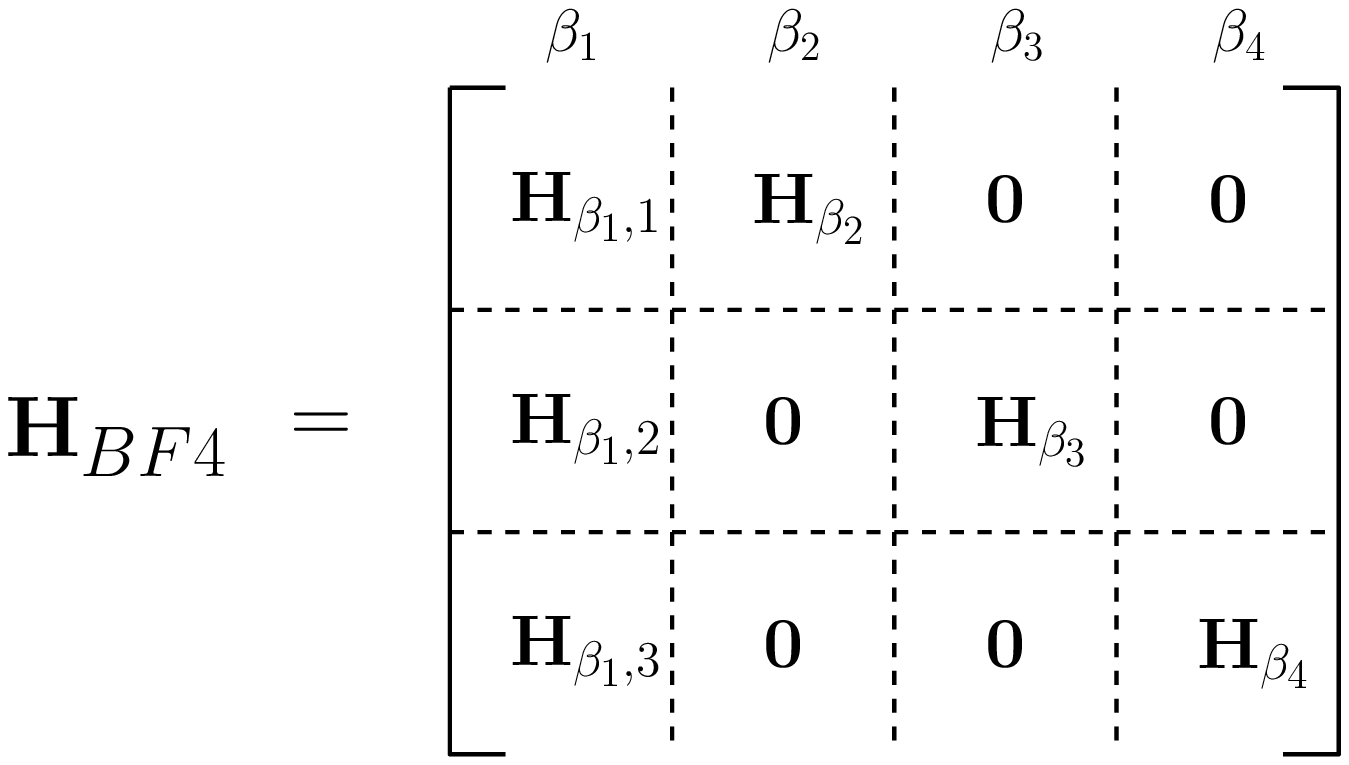}
\caption{Full diversity parity check matrix for the $F=4$ channel}
\label{fig:H_BF4}
\end{figure}

The full diversity code for the $F=4$ channel is provided in Fig. \ref{fig:H_BF4}. Diversity-achieving codes for block fading channels with a greater number of fading channels are constructed in a similar progression as that from the $F=3$ code to the $F=4$ code.

\subsection{Pseudocode for the Proposed Codes}
\label{sec:BF_unstruct_gain}

The pseudocode for construction of the proposed diversity-achieving codes with an arbitrary number of fades, $F$, is provided in Algorithm \ref{alg:BF_unstruct}, demonstrating clearly the separate construction of the submatrices by the PEG-based construction.

\subsection{Rate and Fade Compatible Puncturing}
\label{sec:punct}

From the code graph structures in Figs. \ref{fig:H_BF3} and \ref{fig:H_BF4} for diversity achieving codes on block fading channels with $F=3$ and $F=4$, respectively, we can see that the graph for the $F-1$ channel is effectively nested within the graph for the channel with $F$ fading coefficients. In addition, the graphs are designed to recover from the worst-case scenario of $\alpha_i = 0, i \in \{1,\cdots,F\}$. This allows the use of the graph designed for the channel with $F$ fading coefficients on the $F-1$ channel by means of the elementary puncturing scheme wherein the bits of $\mathbf{V}_F$ are punctured. In this case, only the bits $[\mathbf{V}_1, \mathbf{V}_2, \cdots, \mathbf{V}_{F-1}]$ are transmitted over the block fading channel with $F-1$ fading coefficients. At the input to the decoder, the LLRs associated with the variable nodes in $\mathbf{V}_F$ are set to zero, and iterative decoding is carried out on the full graph for the $F$-channel code. As this is equivalent to an erasure, the properties of the graph ensure that this does not affect the diversity achieving capabilities, with respect to the error rate of the systematic bits.


\section{Simulation Results}
\label{sec:results_Ch4}

The simulation study in this section is presented in three parts. In
the first, the performance results for the unstructured LDPC code
are provided on the binary erasure channel. This demonstrates of the
success of the proposed construction at avoidance of stopping sets
in the graph, as every error event under iterative decoding of LDPC
codes on the BEC is caused by a stopping set \cite{Di_stoppingset}.
The second section provides performance results for the structured
code classes on both BEC and AWGN channels. The results for the AWGN
channel allow easy comparison of performance with the literature. In
the final part of this section, the reduced structure
diversity-achieving codes are evaluated on the block fading channel.
In this case, the results are provided as the variation of the frame
error rate (FER) of the systematic part of the decoded code word as
the channel SNR varies. This is due to the challenging nature of the
channel, meaning that the parity part of the code word is generally
not corrected. This is in contrast to the results provided for the
binary erasure and AWGN channels, where the error rate is computed
for the whole code word as is standard in the literature. The
decoder is operated to a maximum of $40$ decoder iterations for both
BEC and AWGN channels, while a note is made about the choice of
decoder iterations in Section \ref{sec:results_BF}.

For both the general ensemble codes, the QC-LDPC codes and IRA codes, the irregular degree distribution was the density evolution optimised maximum degree $8$ variable node distribution available in the literature \cite{DE1}, Table II:
{
\begin{equation}
\lambda(x) = .30013x + .28395x^2 + .41592x^7
\label{eqn:lambda}
\end{equation}
}
\noindent For all codes constructed, the check node distribution was not specified in the construction algorithm, but rather was forced to have near-regular concentrated form:
{
\begin{equation}
\rho(x) = ax^b + (1-a)x^{b-1}.
\end{equation}
}
Following \cite{RA_Yang}, the variable node degree distribution was constrained to ensure ensure no stopping sets were formed of weight-2 variable nodes alone, a particularly harmful case. As this requirement was applied to all the codes constructed, it does not affect the comparison of construction algorithms presented.

\subsection{Unstructured LDPC Codes}

\begin{figure}[!ht]
\centering
\includegraphics[width=0.4\textwidth]{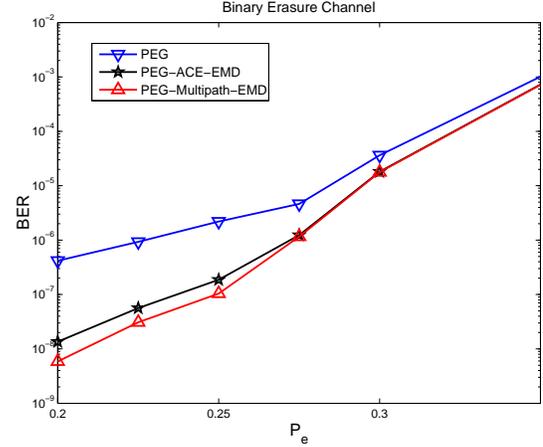}
\caption{Performance comparison on the BEC of the graph construction algorithms for the general LDPC code. The codes are rate $R = \frac{1}{2}$ and block length $N=250$.}
\label{fig:res_BEC2_propcode_prevcode}
\end{figure}

Fig. \ref{fig:res_BEC2_propcode_prevcode} shows the performance of graphs constructed by the construction algorithms considered in Section \ref{sec:metric} on the BEC. As previously stated, these results are significant as they directly demonstrate the relative presence of harmful stopping sets in the respective graphs. The performance improvements in the error floor region shown in Fig. \ref{fig:res_BEC2_propcode_prevcode} are significant.

\subsection{QC-LDPC and IRA Codes}
\label{sec:results_MEMD}

In this section we present results demonstrating the gain achieved through the use of the proposed novel construction algorithm, comparing the short block length performance of a number of classes of codes to those codes constructed by previous methods, the original PEG algorithm \cite{PEG_Hu} and the ACE-based IPEG improvement \cite{Xiao}, along with an algebraic construction for the QC-LDPC codes \cite{QC_Sidon}. The QC-LDPC codes were constructed as in \cite{PEGQC_Li}, with circulant size $8$ which is relatively small compared to the final graph size but is line with the results of that paper. The final distribution of the QC-LDPC codes was thus altered slightly from (\ref{eqn:lambda}) in order to achieve the necessary structure. The algebraic construction based on Sidon sequences was also included in the comparison, in order to provide a point of reference for the performance achieved by the codes constructed. Note that this algebraic construction uses larger tile sizes and therefore achieves greater complexity reduction and possible parallelisation. However, this construction lacks the flexibility of the PEG-based construction algorithm. The IRA codes were constructed by the PEG-based algorithms directly, with the only necessary alteration being the initialisation of the graph associated with the parity bits of the code word to the pre-determined dual diagonal structure of the accumulator.

\begin{figure}[!ht]
\centering
\includegraphics[width=0.4\textwidth]{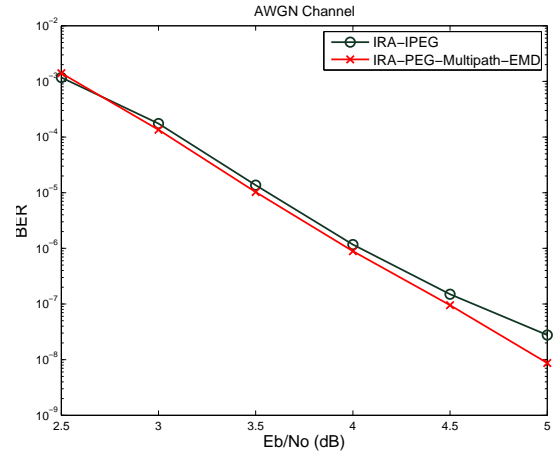}
\caption{Performance of the IRA code graph constructed by the proposed algorithm compared to that of the graph constructed by IPEG algorithm.The codes have rate $R = \frac{1}{2}$ and block length $N=250$.}
\label{fig:res_IRA}
\end{figure}

Improved performance is seen in the error floor region for both the QC-LDPC and IRA codes constructed by the proposed Multipath EMD PEG-based algorithm compared with both the IPEG-based constructions using the ACE metric \cite{Xiao} and the original PEG-based constructions \cite{PEG_Hu}. Fig. \ref{fig:res_IRA} provides the error rate plot for the IRA code class, with the modified IPEG design \cite{Xiao} and the proposed Multipath EMD strategy included on the plot. The IRA graphs constructed have block length $250$ and rate $\frac{1}{2}$.

\begin{figure}[!ht]
\centering
\includegraphics[width=0.4\textwidth]{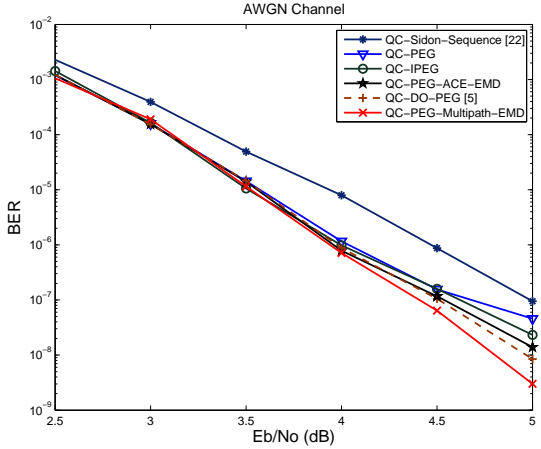}
\caption{Performance of QC-structured LDPC codes of the with rate $R = \frac{1}{2}$ and block length $N=256$ and circulant size $Q=8$.}
\label{fig:res_QC}
\end{figure}

Fig. \ref{fig:res_QC} presents the results for the QC-LDPC codes, with the original PEG \cite{PEG_Hu}, the modified IPEG \cite{Xiao} and the proposed Multipath EMD constructions all used to construct the constrained QC-LDPC irregular code graph \cite{PEGQC_Li} with block length $256$ and submatrix size $Q=8$. The block length of graph constructed using Sidon sequences is $258$, the circulant size is $Q=43$ and the graph is $(3,6)$ regular. Fig. \ref{fig:res_QC} shows that the PEG-based designs provide significant performance improvements over the algebraic construction generally, while the IPEG design offers modest improvements over the PEG construction in the error floor region. The proposed Multipath EMD strategy achieves a gain of $0.4$dB over the PEG construction and $0.3$dB over the IPEG construction at an error rate below $10^{-7}$. Fig. \ref{fig:res_QC} includes results for the previously presented alteration to the IPEG algorithm which makes use of a precise EMD value after the ACE-based decision has been made \cite{PEG_EMD_prev}. It is clear that the strategy presented in this paper outperforms that design in the error floor region. Also included is the plot for the QC-LDPC graph constructed by the QC-DOPEG algorithm \cite{DOPEG_Comms}, demonstrating that although that graph construction offers better performance than the considered constructions from the literature, the method proposed in this work offers best performance overall, with a gain observed of approximately $0.2$dB over the QC-DOPEG constructed graph. This may be accounted for by the fact that the decoder optimisation (DO) operation is applied for a limited number of frames and iterations due to complexity constraints, and that the selection of the noise parameter for testing in the DO operation may be imperfect. The proposed strategy achieves a gain of $0.25$dB over the existing strategy at an error rate below $10^{-7}$.

\begin{figure}[!ht]
\centering
\includegraphics[width=0.4\textwidth]{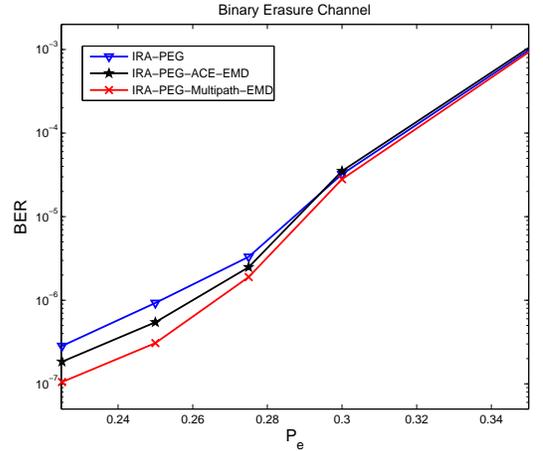}
\caption{Performance comparison on the BEC for the graph construction algorithms applied to the construction of IRA code graphs. The codes are $R = \frac{1}{2}$ and block length $N=230$.}
\label{fig:res_BEC_IRA_results}
\end{figure}

\subsection{Results for the Block Fading Channel}

\label{sec:results_BF}

\begin{figure}[!ht]
\centering
\includegraphics[width=0.4\textwidth]{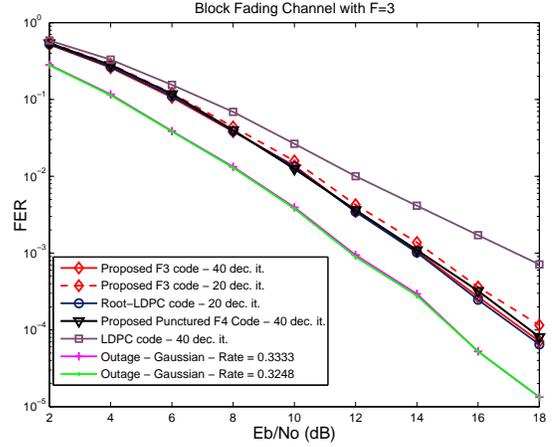}
\caption{Results for the proposed reduced-structure diversity-achieving code on the block fading channel with $F=3$. Included are the Root-LDPC code for that channel and the standard LDPC code of the same dimensions. The plot for the unstructured code designed for the $F=4$ channel and punctured for use on the $F=3$ channel is also included. All graphs are constructed by the PEG algorithm.}
\label{fig:BF_F3_SNR_withPunct}
\end{figure}

\begin{figure}[!ht]
\centering
\includegraphics[width=0.4\textwidth]{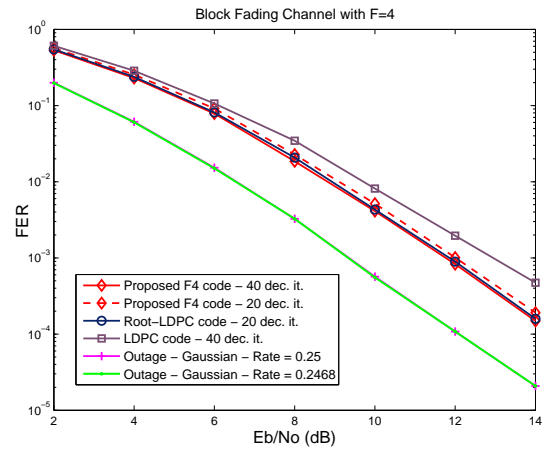}
\caption{Results for the proposed unstructured code for the block fading channel with $F=4$ compared to the Root-LDPC code for that channel and the standard LDPC code. All graphs are constructed by the PEG algorithm.}
\label{fig:BF_F4_SNR}
\end{figure}

\begin{figure}[!ht]
\centering
\includegraphics[width=0.4\textwidth]{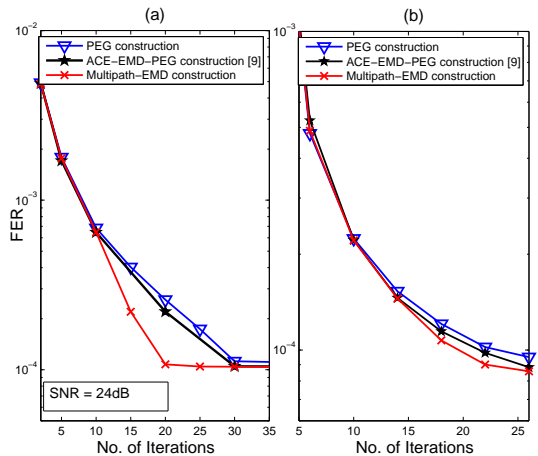}
\caption{Error rate performance against decoder iterations for the codes on the BF channel with $F=2$ and $F=3$, respectively. In (a) the code rate is 0.48, block length $N=248$ and SNR is 24dB while in (b) the code rate 0.3262, block length is $N=282$ and SNR is 18dB. For both, FER is plotted against decoder iteration number.}
\label{fig:BF_F2_F3_iters}
\end{figure}

Simulation results for the block fading channel are presented in Figs. \ref{fig:BF_F3_SNR_withPunct} to \ref{fig:BF_F2_F3_iters}. All codes are irregular, and for channels $F>2$the distributions are derived by density evolution for the AWGN channel, as optimisation \cite{Duyck_BF2_unstruct} remains an open problem for cases with $F>2$. The suboptimal distributions suffice to show that the proposed codes achieve full diversity and  perform close to the Root-LDPC codes with the same distributions. For the $F=2$ unstructured code the degree distribution termed \emph{Code 3} in \cite{Duyck_BF2_unstruct} was used. For all results in this section, the fading coefficients are Rayleigh distributed with scale parameter of $0.5$.

\begin{figure}[!ht]
\centering
\includegraphics[width=0.4\textwidth]{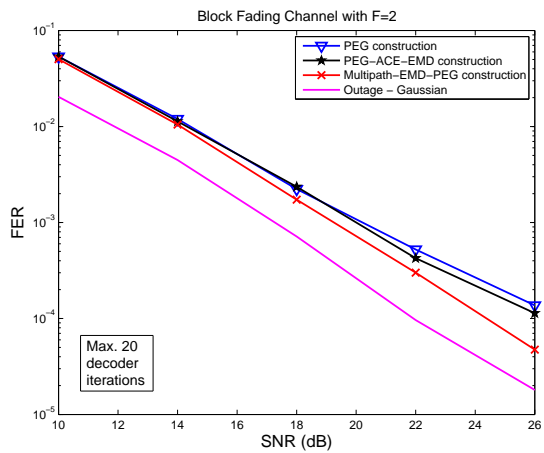}
\caption{A further plot for the $F=2$ code of Fig. \ref{fig:BF_F2_F3_iters}(a) showing the variation of FER with SNR, with the decoder operating to a maximum of 20 iterations.}
\label{fig:BF_F2_SNR}
\end{figure}

Fig. \ref{fig:BF_F3_SNR_withPunct} demonstrates that the proposed code class performs close to the Root-LDPC class, approaching the limit of the channel. However, this figure also demonstrates that the proposed class requires a greater number of iterations ($40$) to converge to the performance of the Root-LDPC graph operating at $20$ decoder iterations. This motivates the results presented in Figs. \ref{fig:BF_F2_F3_iters} and \ref{fig:BF_F2_SNR}. Note that results are also provided for the proposed code graph designed for the $F=4$ channel but punctured for use on the $F=3$ channel, demonstrating this useful feature of the proposed class. Note that for the proposed code class a small rate reduction is imposed in order to meet the requirements of (\ref{eqn:BF_unstruct_requirements}) for achieving full diversity. For the $F=3$ channel, the rate is reduced from $\frac{1}{3}$ to $0.3248$ while for the $F=4$ code of Fig. \ref{fig:BF_F4_SNR} the rate of the proposed code is $0.2468$ rather than $\frac{1}{4}$. All graphs in Fig. \ref{fig:BF_F3_SNR_withPunct} are constructed by the PEG algorithm. Fig. \ref{fig:BF_F4_SNR} demonstrates that the generalisation to higher $F$ is valid.

The slower convergence observed in Figs. \ref{fig:BF_F3_SNR_withPunct} and \ref{fig:BF_F4_SNR} motivated the application of the proposed graph construction to condition the graph. It can be seen clearly in Fig. \ref{fig:BF_F2_F3_iters} that the Multipath EMD PEG construction allows a significantly improved speed of convergence. Fig. \ref{fig:BF_F2_F3_iters} shows the performance of both $F=2$ and $F=3$ codes at fixed SNR over a range of maximum allowed iterations, while Fig. \ref{fig:BF_F2_SNR} shows the performance of the $F=2$ code at a fixed maximum number of iterations of $20$ for a range of SNR values. Both show the performance improvements that the proposed graph construction offer for the new class of codes.

\section{Conclusions}

\label{sec:summary_Ch4}

In this work, a graph-based construction algorithm was proposed
which improves the connection properties of the final graph,
providing performance gains in the error floor region of operation.
The proposed algorithm, called Multipath EMD PEG construction, is
demonstrated to provide significant performance improvements for a
number of useful structured code classes. In addition, a new class
of codes for achieving full diversity on general block fading
channels is presented and is demonstrated to perform competitively
compared to the previously presented code class for this channel.
The novel Multipath EMD construction algorithm is then applied to
the construction of this code class, with improvements in decoder
convergence speed observed as a result. The proposed Multipath EMD
can also be applied to other wireless systems
\cite{spa,mfsic,jiomimo,dfcc,mbdf,did,uchoa2016iterative,tds,gsmi,mbthp,armo,de2013massive,lsmimo}

\appendices

\section{Analysis of the Multipath EMD Metric Progression}

\label{App:MP_EMD}

In a PEG-based construction, any cycles created by placement of the edge $(x_i,v_j)$ will contain that edge, including the shortest-length cycles created. One or many cycles are created when, in Step 6 of Algorithm 1, $\overline{\mathcal{N}_{v_j}^l} = \emptyset$.

In review, the PEG algorithm selects from the set $\mathcal{A} = \{ \overline{ \mathcal{N}_{v_j}^{l-1}}: \overline{ \mathcal{N}_{v_j}^{l}} = \emptyset\}$ the set of nodes with minimum weight, $\mathcal{B} = \{ c_i: |\mathcal{M}_{c_i}^0| = \min\limits_{c_x \in \mathcal{A}} |\mathcal{M}_{c_x}^0|  \}$. In that algorithm, the nodes in this set were considered to be equivalent in terms of their effect on the performance under iterative decoding as they are at equal maximum distance from $v_j$, and so a node was selected from this set at random. In the following, a justification for the decision metric progression employed in the proposed construction algorithm is provided.

Denote the number of shortest-length paths from check node $c_y \in \mathcal{B}$ to the current variable node $v_j$ as $P_{c_y}$ and recall that the set $\mathcal{C} = \{ c_i: P_{c_i} = \min\limits_{c_y \in \mathcal{B}} P_{c_y} \}$. Thus a placement involving any element of $\mathcal{C}$ would create the same minimum number of shortest cycles, $P_{c_i}$. The proposed algorithm selects a node from $\mathcal{C}$ based on the extrinsic connections of those $P_{c_i}$ cycles.

At any particular edge placement in the progressive construction, the original PEG algorithm would create $P_{c_y}$ cycles of length \emph{2l+2}, with $c_y \in \mathcal{B}$ while the Multipath EMD approach of this paper creates $P_{c_i}$ cycles of the same length. By design:
{
\begin{equation}
\label{eqn:proof}
P_{c_i} \leq P_{c_y} ~,~ c_i \in \mathcal{C}, c_y \in \mathcal{B}.
\end{equation}
}
\noindent In the above expression, the equality is satisfied in only two cases, when
{
\begin{equation}
P_{c_y} = P ~ \forall ~ c_y \in \mathcal{B},
\end{equation}
}
\noindent where $P$ is some constant, or when
{
\begin{equation}
|\mathcal{B}| = 1.
\end{equation}
}
\noindent In both of these cases, $\mathcal{C} = \mathcal{B}$. Thus, at worst the proposed metric reduces to that of the original PEG algorithm.

Consider the construction of two code graphs, where all but the final edge placement is made using the same original PEG algorithm. For each placement, the number of shortest cycles created, similarly to the notation used above, $P_{z,L}(\emph{PEG})$ with $z$ indexing the edge placement and $L$ denoting cycle length. Thus the total number of length-4 cycles in the PEG constructed graph is $\displaystyle \sum_{z = 1:E} P_{z,L}(\emph{PEG})$, where $E$ is the total number of edges in the graph. The same applies for cycles of length $L=6,8,\cdots$.

Now, the first graph in our hypothetical situation is constructed entirely by the PEG algorithm, while for the second graph the final placement is made by the proposed Multipath EMD algorithm. In both cases, cycles of length $L = 2\emph{l} + 2$ are created. The total number of cycles of length $2\emph{l} + 2$ in each graph is
{
\begin{equation}
\sum_{z = 1:E} P_{z,2\emph{l} + 2}(\emph{PEG}),
\end{equation}
}
\noindent and
{
\begin{equation}
\sum_{z = 1:E-1} P_{z,2\emph{l} + 2}(\emph{PEG}) + P_{E,2\emph{l} + 2}(\emph{MEMD}),
\end{equation}
}
\noindent respectively.

We wish to show that
{
\begin{equation}
\label{eqn:cycle_ineq}
\sum_{z = 1:E-1} P_{z,2\emph{l} + 2}(\emph{PEG}) + P_{E,2\emph{l} + 2}(\emph{MEMD}) \leq  \sum_{z = 1:E} P_{z,2\emph{l} + 2}(\emph{PEG}),
\end{equation}
}
\noindent Expanding the above equation, we obtain
{
\begin{equation}
\begin{split}
\sum_{z = 1:E-1} P_{z,2\emph{l} + 2}(\emph{PEG}) + P_{E,2\emph{l} + 2}(\emph{MEMD}) \leq \\
 \sum_{z = 1:E-1} P_{z,2\emph{l} + 2}(\emph{PEG}) + P_{E,2\emph{l} + 2}(\emph{PEG}).
\end{split}
\end{equation}
}
\noindent From above, if we assume that $\mathcal{C} \neq \mathcal{B}$,
{
\begin{equation}
P_{E,2\emph{l} + 2}(\emph{PEG}) - P_{E,2\emph{l} + 2}(\emph{MEMD}) = \epsilon,
\end{equation}
}
where $\epsilon$ is some positive integer, while if $\mathcal{C} = \mathcal{B}$,
{
\begin{equation}
P_{E,2\emph{l} + 2}(\emph{PEG}) - P_{E,2\emph{l} + 2}(\emph{MEMD}) = 0,
\end{equation}
}
\noindent proving that (\ref{eqn:cycle_ineq}) holds.

Due to the suboptimal nature of PEG-based constructions, where some choice in edge placement at an earlier stage of the graph, though locally optimal, may negatively impact on available choices for edge placement at a later stage of construction, the corresponding proof may not be constructed for earlier edge placements, $z \le E$. However, the proposed algorithm follows the tractable locally optimal approach of the PEG algorithm and has been demonstrated through simulation to produce graphs capable of excellent performance. As further support for the assertion that reduces the number of shortest length cycles throughout the graph, Table \ref{tab:cyc_dist} provides the total number of cycles of length 6, 8 and 10 in a number of the code graphs used in Fig. \ref{fig:res_QC}, with the cycles counted by means of the algorithm of \cite{lollipop}. Note that the proposed algorithm produces the graph with the fewest number of cycles of length 6 among the constructions considered.

\begin{table*}[t]
  \centering
  \caption{Numbers of cycles of each length found in the code graphs for the graph constructions considered.}

\begin{tabular}{ l || c  c  c }

girth = 6 & QC-PEG & QC-M.-EMD-PEG & QC-PEG-ACE-EMD\\

\hline

No. 6 Cycles & 1560 & 1392 & 1488\\

No. 8 Cycles & 29000 & 30608 & 28888\\

No. 10 Cycles & 462312 & 481320 & 465744\\

\end{tabular}

  \label{tab:cyc_dist}
\end{table*}

\section{On the Complexity of the Proposed Graph Construction Algorithm}

In order to provide an insight into the complexity required for the proposed algorithm in comparison to that of the algorithm in \cite{PEG_EMD_prev}, the number of paths which must be evaluated in order to compute the final EMD metric of each algorithm is considered. To justify this, consider that both algorithms carry out the trivial task of EMD computation once the paths have been identified, while the criterion applied prior to the EMD in the proposed algorithm is based simply on the number of paths identified for each candidate check node.

\begin{figure}[!ht]
\centering
\includegraphics[width=0.4\textwidth]{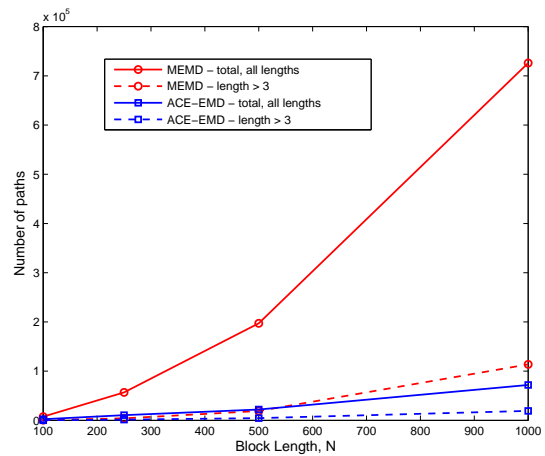}
\caption{A plot of the number of paths computed for the PEG-Multipath-EMD and PEG-ACE-EMD algorithms, respectively, for various block lengths.}
\label{fig:Algs_complexity_paths}
\end{figure}

Fig. \ref{fig:Algs_complexity_paths} provides the total number of paths which must be identified for the algorithms considered, along with the number of paths of length $4$ or longer, leading to cycles of length $10$ or longer. From this figure it is clear that the proposed candidate selection criteria incur a significant cost in terms of complexity. However, note that the majority of path identification operations for the proposed algorithm happen for shorter paths, mitigating somewhat the complexity cost.

\label{App:BF}
\bibliographystyle{IEEEtran}
\bibliography{IEEEabrv,Trans_bib_2}

\end{document}